\documentclass[10pt,twoside,twocolumn,english,aps,prl,superscriptaddress,notitlepage]{revtex4-2}
\usepackage[utf8]{inputenc}
\usepackage[a4paper]{geometry}
\usepackage{color,xcolor,soul}
\usepackage{verbatim,textcomp,enumitem}
\usepackage{amstext,amsfonts,amsmath,amssymb}
\usepackage{graphicx}
\usepackage[calcwidth,explicit]{titlesec}
\usepackage{chngcntr}
\usepackage[normalem]{ulem}
\usepackage{soul}
\usepackage{amsmath}
\usepackage[textsize=scriptsize,textwidth=2cm]{todonotes}
\usepackage[unicode=true, pdfusetitle, bookmarks=false, breaklinks=true, pdfborder={0 0 0}, pdfborderstyle={}, backref=false, colorlinks=true]{hyperref}
\hypersetup{colorlinks=true, citecolor=blue, linkcolor=blue, urlcolor=blue}

\makeatletter\renewcommand\frontmatter@abstractwidth{\dimexpr\textwidth-2cm\relax}\makeatother

\titleformat{\section}{}{}{0pt}{}
\titleformat{\section}{\bfseries\sffamily\filcenter}{}{0.2em}{#1}
\titlespacing{\section}{0pt}{0.2ex}{0.1ex}
\titleformat{\paragraph}[runin]{\normalfont\normalsize\bfseries}{}{0pt}{\theparagraph.}
\titlespacing*{\paragraph}{0em}{0ex}{0.3em}[]

{}

\setcitestyle{super}

\counterwithout{paragraph}{subsubsection}
\renewcommand{\theparagraph}{\arabic{paragraph}}

\AtBeginDocument{
	\renewcommand{\ref}[1]{\autoref{#1}}

}

\begin{document}
\title{Stability and Character of Zero Field Skyrmionic States \\in Hybrid Magnetic Multilayer Nanodots\smallskip{}}

\author{Alexander Kang-Jun Toh}
\affiliation{Institute of Materials Research \& Engineering, Agency for Science, Technology \& Research, 138634 Singapore}
\author{McCoy W. Lim}
\affiliation{Department of Physics, National University of Singapore, 117551 Singapore}
\author{T. S. Suraj}
\affiliation{Department of Physics, National University of Singapore, 117551 Singapore}
\author{Xiaoye Chen}
\affiliation{Institute of Materials Research \& Engineering, Agency for Science, Technology \& Research, 138634 Singapore}
\author{Hang Khume Tan}
\affiliation{Institute of Materials Research \& Engineering, Agency for Science, Technology \& Research, 138634 Singapore}
\author{Royston Lim}
\affiliation{Institute of Materials Research \& Engineering, Agency for Science, Technology \& Research, 138634 Singapore}
\author{Xuan Min Cheng}
\affiliation{Department of Materials Science and Engineering, National University of Singapore, 117575 Singapore}
\author{Nelson Lim}
\affiliation{Institute of Materials Research \& Engineering, Agency for Science, Technology \& Research, 138634 Singapore}
\author{Sherry Yap}
\affiliation{Institute of Materials Research \& Engineering, Agency for Science, Technology \& Research, 138634 Singapore}
\author{Durgesh Kumar}
\affiliation{Division of Physics and Applied Physics, School of Physical and Mathematical Sciences, Nanyang Technological University, 637371 Singapore}
\author{S. N. Piramanayagam}
\affiliation{Division of Physics and Applied Physics, School of Physical and Mathematical Sciences, Nanyang Technological University, 637371 Singapore}
\author{Pin Ho}
\email{hopin@imre.a-star.edu.sg}
\affiliation{Institute of Materials Research \& Engineering, Agency for Science, Technology \& Research, 138634 Singapore}
\author{Anjan Soumyanarayanan}
\email{anjan@nus.edu.sg}
\affiliation{Department of Physics, National University of Singapore, 117551 Singapore}
\affiliation{Institute of Materials Research \& Engineering, Agency for Science, Technology \& Research, 138634 Singapore}

\begin{abstract}
    Ambient magnetic skyrmions stabilized in multilayer nanostructures are of immense interest due to their relevance to magnetic tunnel junction (MTJ) devices for memory and unconventional computing applications. 
    However, existing skyrmionic nanostructures built using conventional metallic or oxide multilayer nanodots are unable to concurrently fulfill the requirements of nanoscale skyrmion stability and feasibility of all-electrical readout and manipulation. 
    Here, we develop a few-repeat hybrid multilayer platform consisting of metallic [Pt/CoB/Ir]$_3$ and oxide [Pt/CoB/MgO] components that are coupled to evolve together as a single, composite stack. 
    Zero-field (ZF) skyrmions with sizes as small as 50 nm are stabilized in the hybrid multilayer nanodots, which are smoothly modulated by up to 2.5$\times$ by varying CoB thickness and dot sizes.
    Meanwhile, skyrmion multiplets are also stabilized by small bias fields. 
    Crucially, we observe higher order "target" skyrmions with varying magnetization rotations in moderately-sized, low-anisotropy nanodots. 
    These results provide a viable route to realize long-sought skyrmionic MTJ devices and new possibilities for multi-state skyrmionic device concepts.
    
\end{abstract}
\maketitle


\section{Introduction}
\paragraph{Motivation for Dots}
    Magnetic skyrmions are nanoscale spin textures with fixed handedness, or chirality \cite{Nagaosa.2013, Fert.2017}. 
    They can be stabilized at room temperature in commercially viable multilayered magnetic thin films \cite{ Boulle.2016, Moreau-Luchaire.2016,  Soumyanarayanan.2017}, wherein they  exhibit efficient current-induced motion \cite{Woo.2016}, nucleation \cite{Jiang.2015}, and topological dynamics \cite{Litzius.2017, Jiang.2017}. 
    These device attributes are promising for applications in next-generation memory \cite{Bhattacharya.2018, Hagemeister.2015, Tey.2022}, logic \cite{Zhang.2015, Luo.2018, Zazvorka.2019}, and computing technologies \cite{Torrejon.2017, Grollier.2020, Song.2020}. 
    The stability and character of chiral spin textures are governed by the interplay of conventional and chiral magnetic interactions, with the latter arising from heavy metal (HM) $-$ ferromagnet (FM) interfaces \cite{Wiesendanger.2016, Soumyanarayanan.2017, Xiaoye.2022}. 
    Within nanoscale device geometries, confinement effects considerably influence their formation and evolution\cite{Boulle.2016, Finocchio.2016, Ho.2019}.   
    A comprehensive picture of skyrmion stability in ultra-thin nanodots is especially crucial for robust realizations of highly sought-after magnetic tunnel junctions (MTJs) towards all-electrical readout and manipulation \cite{Fert.2017, Back.2020}.
   
\paragraph{Literature Review}
    Early works on confining nanoscale skyrmions predominantly utilized numerous stacked repetitions of metallic multilayers ([HM/FM/HM]$_N$, $N > 10$) \cite{Moreau-Luchaire.2016, Woo.2016, Zeissler.2017, Ho.2019}, wherein texture stability is aided considerably by interlayer dipolar interactions \cite{Legrand.2018}.  
    In contrast, MTJ compatibility requires an ultrathin metallic multilayer, capped with a single oxide layer \cite{Back.2020}. 
    However, chiral nanodots comprising a single oxide trilayer (HM/FM/oxide) nucleate skyrmions only over a narrow window of magnetic and geometric parameters \cite{Boulle.2016, Juge.2018}, with limited utility for scalable MTJ development. 
    In contrast, for few-repeat ($N\leq 3$) metallic multilayers, spin texture stability is considerably enhanced through interlayer exchange coupling (IEC) \cite{LoConte.2020, Xiaoye.2023}, with the added ability to generate strong spin-orbit torques for efficient manipulation \cite{Finizio.2019, Zeissler.2020}.
    This motivates case for nanostructures comprising metallic and oxide multilayers, to realize stable skyrmions with electrical detection, and efficient dynamics. 
    
\paragraph{Results Summary}
    In this work, we develop few-repeat hybrid chiral multilayer nanodots by combining metallic (Pt/CoB/Ir) and oxide (Pt/CoB/MgO) stacks, and engineering them to evolve together. 
    Using magnetometry, magnetic force microscopy (MFM), and micromagnetic simulations, we show that such nanodots behave as one composite stack, with chiral spin textures stabilized over a wide range of FM thicknesses and dot sizes, including at ZF. 
    The spin textures exhibit a variety of configurations, including single, multiple and higher-order skyrmions, and exhibit marked variation with magnetic parameters, geometry, and external field.
    This provides a strategic platform to integrate the benefits of metallic and oxide multilayer stacks, with immediate relevance to functional spintronic devices.

\begin{figure} [!htbp]
    \includegraphics[width = \linewidth]{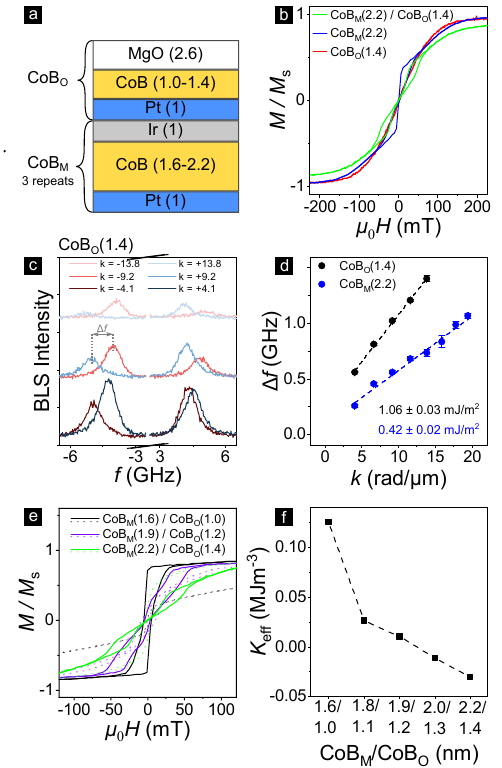}
        \caption{
        \textbf{Magnetic properties of hybrid multilayer stack.} 
        \textbf{(a)} Schematic of hybrid multilayer stack comprising [Pt/CoB/Ir]$_3$ and [Pt/CoB/MgO], represented as CoB$_{\rm M}$ and CoB$_{\rm O}$, respectively (layer thickness in nm in parentheses). 
        \textbf{(b)} Out-of-plane (OP) magnetization hysteresis loops of CoB$_{\rm M}$(2.2), CoB$_{\rm O}$(1.4), and CoB$_{\rm M}$(2.2)/CoB$_{\rm O}$(1.4). The CoB thicknesses of CoB$_{\rm M}$ and CoB$_{\rm O}$ were chosen to have similar saturation fields ($H_{\rm s}$). 
        \textbf{(c)} Representative Brillouin light scattering (BLS) of CoB$_{\rm O}$(1.4) for several wave vectors ($k$), showing Stokes and anti-Stokes peaks for applied fields of +500 mT (blue) and -500 mT (red). 
        \textbf{(d)} Dispersion of BLS peak frequency shift, $\Delta f$, between counter propagating spin waves for CoB$_{\rm O}$(1.4) (black) and CoB$_{\rm M}$(2.2) (blue). Dashed lines indicate linear fits, used to determine the respective interfacial Dzyaloshinskii–Moriya interactions (iDMIs). 
        \textbf{(e)} OP (solid) and in-plane (IP, dotted) hysteresis loops for three of the five CoB$_{\rm M}$(1.6--2.2)/CoB$_{\rm O}$(1.0--1.4) stacks, encompassing the range of magnetization evolution in this work. 
        \textbf{(f)} Measured effective anisotropy ($K_{\rm eff}$) of the five hybrid chiral multilayers studied in the work.}
    \label{fig:MagChar}
\end{figure}

\section{Hybrid Multilayer Structure}
\paragraph{Methods}
    Double wedged multilayers with active stack composition [Pt(1)/CoB(1.6--2.2) /Ir(1)]$_3$/[Pt(1)/CoB(1.0--1.4)/MgO(2.6)] (thickness in nm in parentheses) were sputter-deposited onto 8" Si/SiO$_2$ wafers using the Singulus Timaris$^\text{TM}$ ultra-high vacuum magnetron sputtering tool (see SI §S1).
    The films were patterned into dots with diameters, $W$ of 300--1000 nm using electron beam lithography (Elionix EX10$^\text{TM}$) and ion beam etching (Oxford CAIBE$^\text{TM}$) (see SI §S2).
    Magnetization hysteresis loops were measured using the MicroMag Model 2900$^\text{TM}$ alternating gradient magnetometer (see SI §S1).
    The chirality, manifested as the interfacial Dzyaloshinskii-Moriya interaction (iDMI), was measured using Brillouin light scattering (BLS) spectroscopy, performed on a wave vector resolved set-up with a six-pass tandem Fabry-Perot interferometer (see SI §S1).
    Spin textures were imaged by MFM performed using the Bruker Dimension ICON$^\text{TM}$ microscope with Nanosensors SSS-MFMR$^\text{TM}$ tips (diameter $<$ 30 nm, ultra-low magnetization $\sim$80 emu/cm$^3$), including with \textit{in situ} OP magnetic fields (0--60 mT) (see SI §S3).
    Finally, to interpret the experiments, grain-free micromagnetic simulations using the effective medium approximation \cite{Woo.2016} were performed with mumax$^3$ program \cite{Vansteenkiste.2014}, using magnetic parameters, stack geometry, and dot sizes consistent with experiments (see SI §S4). 

\paragraph{Hybrid Stack Structure}
    Our multilayer films (schematic: \ref{fig:MagChar}(a)) combine two distinct chiral stacks – a metallic stack [Pt/CoB$(x)$/Ir]$_3$ (represented as CoB$_{\rm M}(x)$) and oxide component [Pt/CoB$(y)$/MgO] (CoB$_{\rm O}(y)$).
    The 3-repeat chiral CoB$_{\rm M}$ stack was shown to host skyrmions with tunable properties at film-level \cite{HK.2021, Lucassen.2019}, and enables their spin-orbit torque driven nucleation and dynamics in devices \cite{ Finizio.2019, Zeissler.2020}. 
    Meanwhile, the CoB$_{\rm O}$ multilayer is capped with MgO, which can serve as a tunnel barrier for electrical detection of chiral spin textures \cite{Romming.2013, Hanneken.2015, Li.2022, Guang.2022}.
    The two component stacks are integrated by a metallic Ir(1)/Pt(1) spacer, which is known to exhibit strong IEC \cite{Karayev.2019, Xiaoye.2023}. 

\paragraph{Magnetic Properties of Stack Components}
    The stack components were carefully engineered to ensure their evolution as one chiral multilayer. 
    First, magnetometry measurements were performed to determine the out-of-plane (OP) saturation fields, $H_{\rm s}$, of the individual CoB$_{\rm M}$ and CoB$_{\rm O}$, in each case, over a range of CoB thicknesses  (CoB$_{\rm M}$: 1.6--2.2 nm; CoB$_{\rm O}$: 1.0--1.4 nm). 
    These measurements were used to calibrate and match pairs of CoB$_{\rm M}$ and CoB$_{\rm O}$ stacks with similar $H_{\rm s}$ values (see SI §S1). 
    For example, in \ref{fig:MagChar}(b), the $H_{\rm s}$ for CoB$_{\rm M}$(2.2) (blue) and CoB$_{\rm O}$(1.4) (red) is  $\sim$180 mT.  
    Finally, the double-wedged composite stack was fabricated using matched CoB$_{\rm M}$ and CoB$_{\rm O}$ thicknesses to ensure coherent switching of the hybrid multilayer.
    \ref{fig:MagChar}(b) shows the resulting OP magnetization hysteresis loop, $M(H)$, of the CoB$_{\rm M}$(2.2)/CoB$_{\rm O}$(1.4) stack (green), which exhibits the distinctive shear characteristic of chiral multilayers. 
    Meanwhile, wave-vector resolved BLS spectroscopy measurements, shown in \ref{fig:MagChar}(c) for CoB$_{\rm M}$(2.2) and CoB$_{\rm O}$(1.4), establish the left-handed N\'eel chirality of both stack components. 
    The linear dispersion of the counter propagating spin-waves (\ref{fig:MagChar}(d)) was used to determine the iDMI, $D$ \cite{Kuepferling.2023}, found to be 0.42\,mJ/m$^2$ (for CoB$_{\rm M}$(2.2)) and 1.06 \,mJ/m$^2$ (for CoB$_{\rm O}$(1.4)), respectively.
    The magnitude and sign of the iDMI values are in line with published reports on similar metallic \cite{Alshammari.2021, Finizio.2019b} and oxide \cite{Ma.2018b, Bhatti.2023} multilayers, and are sufficiently large to stabilize homochiral spin textures \cite{Xiaoye.2023}.
    
\begin{figure*} [!htbp]
    \includegraphics[width = \linewidth]
    {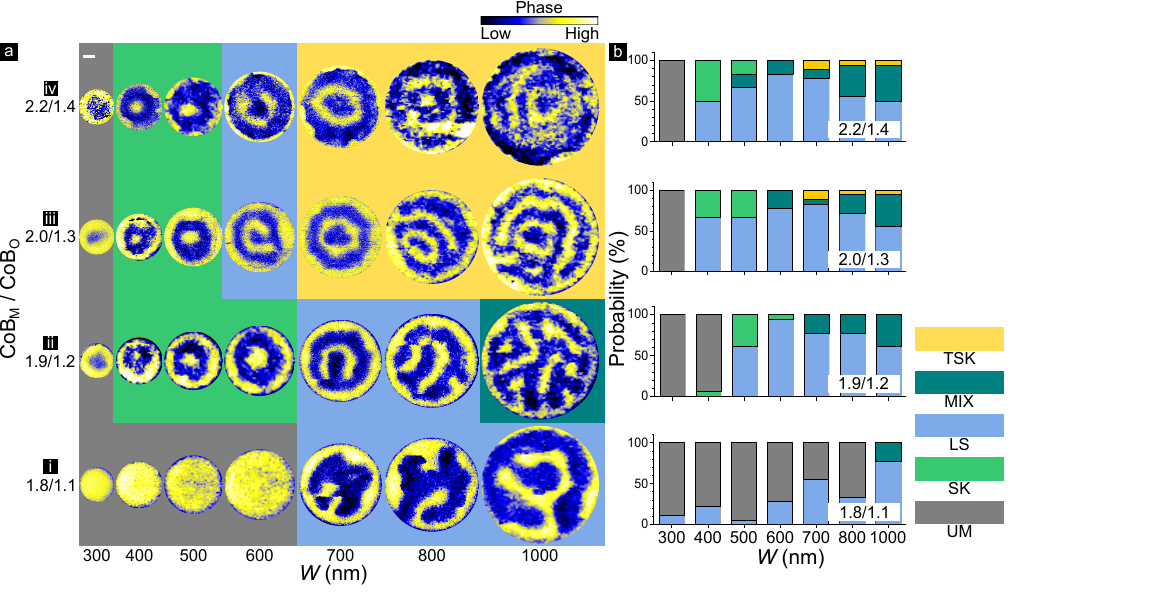}
        \caption{
        \textbf{Confined Zero-field (ZF) Spin Textures.} 
        \textbf{(a)} Magnetic force microscopy (MFM) images (scalebar: 100 nm) at ZF of spin texture configurations in nanodots of varying sizes ($W$, left-to-right), and for varying multilayer compositions (i-to-iv).
        Background color indicates identified state of dot: uniform magnetization (UM, grey), skyrmion(s) (SK, green), labyrinthine stripes (LS, blue), mixed LS, SK (MIX, teal), and target skyrmion (TSK, yellow), respectively. 
        \textbf{(b)} Empirically determined ZF probability  of stabilizing the respective states for varying multilayers and dot sizes, determined by averaging over 18 nominally identical dots for each presented dataset.
        }
    \label{fig:MFM-ZF}
\end{figure*}

\paragraph{Evolution of Magnetic Properties}
    The stability and characteristics of chiral spin textures in multilayers are determined by the interplay of symmetric exchange, $A$, iDMI, and effective anisotropy, $K_{\rm eff}$ \cite{Soumyanarayanan.2017, Xiaoye.2022}, each of which may vary across the wedged multilayer. 
    For this work, $D$, quantified for the respective stack components in \ref{fig:MagChar}(d), is treated as approximately constant across the wedge, as the CoB thickness varies by  $< 30$\% for both stack components \cite{Cho.2015}.
    Similarly, $A$ is also expected to vary negligibly, given the $<10$\% variation in the measured saturation magnetization, $M_{\rm s}$  across CoB thicknesses \cite{Nembach.2015}.
    In contrast, the $K_{\rm eff}$ should be readily tuned with varying CoB thickness \cite{HK.2021}. 
    Indeed, the $M(H)$ loops in \ref{fig:MagChar}(e) show that with increasing CoB thickness, the easy axis of the hybrid stack transitions from OP to in-plane (IP) orientation. The resulting $K_{\rm eff}$ is modulated over -0.03--0.126 MJ/m$^3$ (\ref{fig:MagChar}(f)), and the OP-IP crossover  ($K_{\rm eff} \sim 0$) occurs between samples CoB$_{\rm M}$(1.9)/CoB$_{\rm O}$(1.2) and CoB$_{\rm M}$(2.0)/CoB$_{\rm O}$(1.3).

\section{Skyrmion Stability}
\paragraph{Zero-field MFM}
    \ref{fig:MFM-ZF}(a) shows representative ZF MFM images of our nanodots across varying CoB$_{\rm M}$(1.8--2.2)/CoB$_{\rm O}$(1.1--1.4) thicknesses and $W$ over 300—1000 nm. 
    The nanodots exhibited five distinct magnetic states at ZF, identified as uniform magnetization (UM), skyrmion(s) (SK), labyrinthine stripes (LS), mixed-state (MIX), and target skyrmions (TSK).
    While the first three states (UM, SK, LS) are well-established \cite{Woo.2016, Ho.2019}, a mixed state reflects the presence of both SK and LS states.  Meanwhile, target skyrmions are higher-order skyrmions that appear as annular or concentric domains, discussed in detail subsequently  \cite{Leonov.2013, Zheng.2017}. 
    As seen in \ref{fig:MFM-ZF}(a), these ZF magnetic states evolve systematically with varying CoB thickness, and $W$.  
    First, for the thinnest CoB sample -- CoB$_{\rm M}$(1.6)/CoB$_{\rm O}$(1.0) (not shown) the nanodots display UM states across all $W$, with the preferential stability of single domains arising from the substantially larger film-level $K_{\rm eff}$ (\ref{fig:MagChar}(f)) compared to the other samples.
    Next, for CoB$_{\rm M}$(1.8)/CoB$_{\rm O}$(1.1) nanodots (\ref{fig:MFM-ZF}(a): bottom), reducing $W$ results in a sharp transition from LS to UM states, albeit without any intermediate states.
    As the CoB thickness is further increased across CoB$_{\rm M}$(1.9--2.2)/CoB$_{\rm  O}$(1.2--1.4) (\ref{fig:MFM-ZF}(a): bottom to top), reducing $W$ results in transitions from LS, MIX, or TSK states to the  SK state, observed for $W \sim 400-600$ nm. 
    As $W$ is further reduced, the dot eventually transitions to the UM state at ZF.         

\paragraph{Zero-field Probability}
    To elucidate the ZF stability of skyrmions confined in multilayer nanodots, we note that the relevant tuning parameter, $K_{\rm eff}$, includes contributions from magneto-crystalline anisotropy, shape anisotropy ($K_{\rm s}$), and interface anisotropy ($K_{\rm i}$) \cite{Johnson.1996}. 
    A suitably moderate $K_{\rm eff}$ required for ZF skyrmion stability can be achieved by modulating the counteracting effects of reducing $W$, for larger $K_{\rm s}$, and increasing CoB, for smaller $K_{\rm i}$, respectively.   
    To this end, we statistically examine the stability of the varied magnetic states across compositions and $W$, shown  in \ref{fig:MFM-ZF}(b), obtained by characterizing $\sim$18 nominally identical nanodots for each composition and $W$ (see SI §3). First, for CoB$_{\rm M}$(1.8)/CoB$_{\rm O}$(1.1) (\ref{fig:MFM-ZF}(b): bottom), the high $K_{\rm i}$ is inherently unfavourable for stabilizing skyrmions \cite{Soumyanarayanan.2017, Xiaoye.2022}, and thus the sparse LS or mixed configurations directly transition to a stable UM configuration at a relatively large $W \lesssim 600$ nm.
    As $K_{\rm i}$ is decreased (\ref{fig:MFM-ZF}(b): bottom to top), LS, mixed, and TSK states are stabilized at large $W\gtrsim 700$ nm, with greater preference for mixed over LS states in comparison to film-level magnetic textures.
    For these compositions, the modulation of $K_{\rm s}$ with $W$ enables the stabilization of ZF skyrmions for intermediate $W$ (400--600 nm), wherein the required $K_{\rm eff}$ is realized. Skyrmion nucleation probabilities of up to 50\% are observed in 400-500 nm-sized dots with optimal compositions.
    Notably, the optimal window for ZF skyrmion stability shifts to smaller $W$, and increases in extent with increasing CoB thickness, in line with the expected modulation of $K_{\rm s}$ and $K_{\rm i}$ respectively. 

\paragraph{Target Skyrmions Intro}
    Next, we turn to our striking observation of target skyrmions, which present as concentric domains, stabilized by confined geometries \cite{Leonov.2013}. We define $R_M$ to be the cumulative rotation of the magnetization vector as one traces from the center of the skyrmion radially to infinity. Target skyrmions, by definition, have $R_M > \pi$ \cite{Leonov.2013, Zheng.2017, Kent.2019}, compared to $R_M = \pi$ for a skyrmion \cite{Nagaosa.2013}.  
    Target skyrmions have recently been observed in nanodisks of several crystalline helimagnetic systems known to host Bloch skyrmions \cite{Yu.2010, Zheng.2017, Repicky.2021, Zhang.2023}.
    However, within multilayers, there are no reports of their inherent stability, notwithstanding structural imprinting achieved using underlying vortex structures \cite{Kent.2019}.

\begin{figure} [!htbp]
    \includegraphics [width = \linewidth]{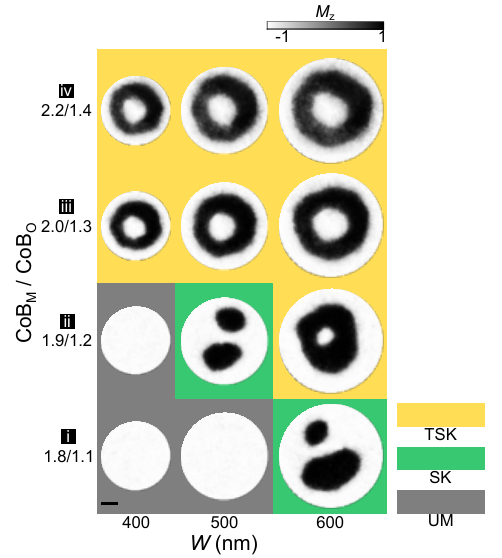}
        \caption{
        \textbf{ Simulated ZF Spin Textures in Nanodots.} 
        Micromagnetic simulations of OP magnetization ($m_z$, scale bar: 100 nm) for nanodots with varying $W$ (left-to-right), and varying multilayer-equivalent effective stack parameters (i-to-iv). 
        Background colour indicates identified state of dot: UM (grey), SK (green), and TSK (yellow), respectively.
        }
    \label{fig:Sims}
\end{figure}
    
\paragraph{Target Skyrmions}
    Here we observe ZF target skyrmions with $R_M$ of 2$\pi$ and $\sim$300 nm sizes inherently stabilized in low-$K_{\rm i}$-nanodots of $W$ $\gtrsim$ 700 nm (\ref{fig:MFM-ZF}(a): top right), and occasionally, with higher-order $R_M$ (3$\pi$, i.e. two concentric rings) as well. 
    The nucleation of TSKs in our hybrid multilayer nanodots, and their absence in previous such works, can be attributed to several key factors: (a) the variation of magnetization between the distinct, albeit coupled CoB$_{\rm M}$ and CoB$_{\rm O}$ stacks \cite{Zheng.2017}, (b) low interfacial anisotropy \cite{Kent.2019}, and (b) the presence of moderate geometric confinement, and (d) few-repeat stack geometry \cite{Rohart.2013}, the confluence of which likely coerce the elongated chiral stripes to coil onto themselves, or induce the nucleation of a smaller skyrmion within an oppositely-polarized skyrmion. 
    While absent in corresponding thin films, such target skyrmions are nucleated with $\sim$10-15\% probability at optimal dot sizes, against LS and mixed states. Their optimal stability at $W \sim 700$ nm dots of thick CoB, suggest that, analogous to conventional skyrmions, their stability can be attributed to a window of $K_{\rm eff}$, arising from moderately large $W$, and the low $K_{\rm i}$ of thick CoB. 
    These conditions may be required to support the in-plane curling critical for the stability of higher-order skyrmions relative to LS and mixed states.

\paragraph{Micromagnetic Simulations}
    To elucidate the interplay of confinement and anisotropy effects on skyrmion stability in our hybrid multilayer nanodots, we performed grain-free micromagnetic simulations using the effective medium approximation.
    The stack structures and magnetic parameters were chosen to emulate their experimentally studied counterparts (CoB$_{\rm M}$(1.6--2.2)/CoB$_{\rm O}$(1.0--1.4)) and the ZF states of nanodots were studied for a range of $W$ of 400--600 nm (\ref{fig:Sims}, see SI §S4).
    As seen in experiments, simulations support the stability of TSKs in nanodots with near-zero or negative $K_{\rm eff}$ (\ref{fig:Sims}: top half).  
    Meanwhile for dots with intermediate $K_{\rm eff}$ (CoB$_{\rm M}$(1.9)/CoB$_{\rm O}$(1.2)-equivalent), we observe a multi-state transition with reducing $W$, from TSK to SK to UM states.  
    Finally, for intermediate values of $W$ and $K_{\rm eff}$, the simulations suggest the stability of skyrmionic multiplets, which have been experimentally observed in nanodots with higher repetitions \cite{Ho.2019}. 
    While the simulations qualitatively reproduce the experimentally observed trends of skyrmion stability evolution with $W$ and $K_{\rm i}$, we note quantitative discrepancies that inhibit one-to-one comparisons with experiment.
    One likely source is the imprecise magnitude of interlayer exchange coupling (IEC), both within the CoB$_{\rm M}$ stack and between the CoB$_{\rm M}$ - CoB$_{\rm O}$.
    In contrast to multilayers with high stack repetitions, where dipolar effects assume greater importance \cite{Legrand.2018}, IEC is known to influence the stability of ZF spin textures in few-repeat multilayers \cite{LoConte.2020, Legrand.2020, Xiaoye.2023}.
    Future simulation works may build on accurately determined IECs, e.g., using bespoke magnetometry techniques\cite{Legrand.2020, Xiaoye.2023, Shaohai.2023}, to quantitatively elucidate transitions between SK and TSK states, towards their practical utilization.

\begin{figure*} [!htbp]
    \includegraphics [width = \linewidth]{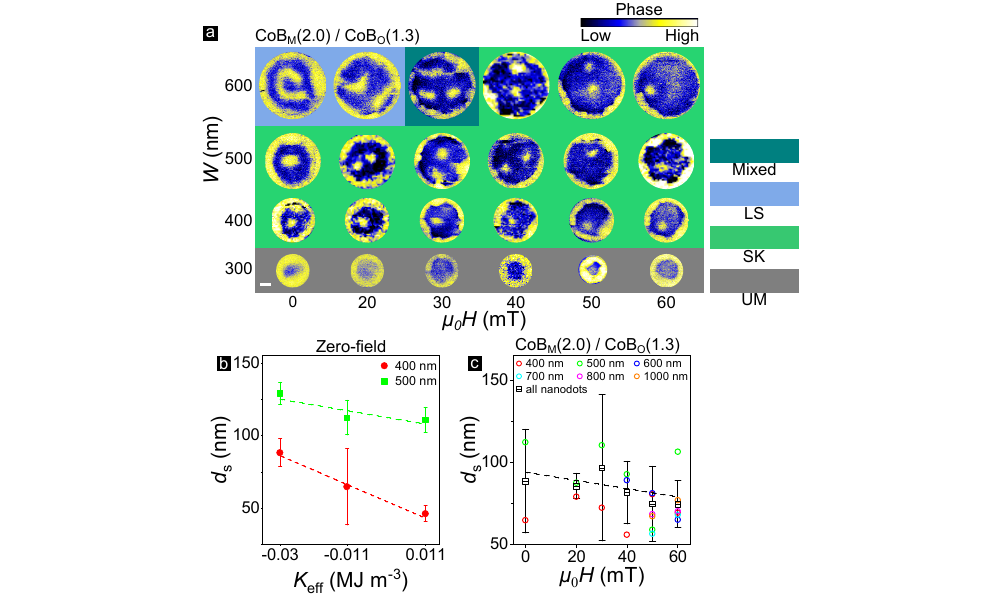}
        \caption{
        \textbf{Magnetic Field Dependence of Confined Skyrmion Properties.} 
        \textbf{(a)} MFM images of CoB$_{\rm M}$(2.0)/CoB$_{\rm O}$(1.3) nanodots (scale bar: 100 nm) for varying \emph{in situ} OP fields ($\mu_0H$, left-to-right, acquired as independent datasets) and with varying $W$ (bottom-to-top).  
        Background color indicates identified state of dot (see \ref{fig:MFM-ZF}(a)).
        \textbf{(b)} Average size of skyrmions ($d_{\rm S}$, from 18 nominally identical nanodots for each size), stabilized at ZF in $W \simeq 400, 500$ nm dots, for varying multilayer compositions ($K_{\rm eff}$ reduces right-to-left). Dashed lines are linear fits. 
        \textbf{(c)} Field-evolution of average $d_{\rm S}$ (from 9 nominally identical nanodots for each size) for sample CoB$_{\rm M}$(2.0)/CoB$_{\rm O}$(1.3) as a function of $\mu_0 H$ for all $W$. Dashed line is a linear fit to the average $d_{\rm S}$ evolution.
        }
    \label{fig:MFM-Size}
\end{figure*}

\section{Skyrmion Size Dependence}
\paragraph{In Field Effect}
    Next, we examine the evolution of skyrmion stability with \emph{in situ} magnetic field ($\mu_0H$).  
    \ref{fig:MFM-Size}(a) shows the MFM-imaged magnetic states for representative CoB$_{\rm M}$(2.0)/CoB$_{\rm O}$(1.3) nanodots as a function  of $W$ and $\mu_0H$. 
    Expectedly, the $W \simeq 300$ nm dots remain uniform throughout, and $W \simeq 400$ nm dots consistently form a single skyrmion over the field range. Meanwhile, for the larger 500 and 600 nm dots, a small OP field is sufficient to destabilize LS and TSK states, giving way to skyrmion multiplets over moderate fields (20--50 mT), which diminish to single skyrmions at higher fields en route to saturation. 
    Finally, the size of skyrmions, $d_{\rm S}$, is also visibly influenced by the applied field, stack properties, and geometry, as detailed below.

\paragraph{ZF Skyrmion Size}
    \ref{fig:MFM-Size}(b) shows the evolution of ZF skyrmion size for dot sizes, $W \simeq 400, 500$ nm, previously shown to host skyrmions across most samples (see \ref{fig:MFM-ZF}, SI §S3). 
    The ZF $d_{\rm S}$ goes as low as $\sim 50$ nm and varies by up to 2.5$\times$ with CoB thicknesses. 
    Expectedly, for larger $W$, the measured $d_{\rm S}$ is greater, and exhibits reduced sensitivity to sample composition \cite{Rohart.2013, Ho.2019}.
    Meanwhile, the linear increase in ZF sizes of isolated skyrmions with decreasing $K_{\rm eff}$ (increasing CoB thickness) is not reported in any other system to the best of our knowledge. 
    In this regard, our prior work on high-repeat multilayer nanodots, which reported a direct correlation between $d_{\rm S}$ and $D/A$ for skyrmion multiplets, also included an implicit direct correlation with $K_{\rm eff}$, and therefore involved a complex interplay of multiple parameters and skyrmion interactions. 
    In contrast, the large (2.5$\times$) tunability of ZF $d_{\rm S}$ achieved here with a single-magnetic knob ($K_{\rm eff}$) offers greater utility towards functional devices.

\paragraph{In Field Skyrmion Size}
    Finally, we examine the field-dependence of $d_{\rm S}$ for the case of moderate $K_{\rm eff}$ (CoB$_{\rm M}$(2.0)/CoB$_{\rm O}$(1.3)) by imaging 9 nominally identical nanodots for each field and $W$ (\ref{fig:MFM-Size}(c)). Expectedly, smaller skyrmions are observed with increasing $\mu_0H$, with the average $d_{\rm S}$ across dots decreasing from $\sim 85$ nm to $\sim 65$ nm.
    While smaller dots ($W \lesssim 500$ nm) hosting individual skyrmions exhibit consistent and sizable $H$-variation, $d_{\rm S}$ is found to be relatively independent of $W$ for larger dots ($W \gtrsim 500$ nm). 
    Consistent with our previous work \cite{Ho.2019} (albeit for different stack parameters), the $W$-independence emerges at these dot sizes as a result of the emergence of skyrmion multiplets at finite fields, due to which the role of geometric confinement effects is diminished. 
    Overall, these results demonstrate that the sub-100 nm ZF skyrmions stabilized in our hybrid multilayer nanodots exhibit consistent field evolution trends driven by stack properties and confinement effects.

\section{Conclusion}
\paragraph{Summary}
    In summary, we report on chiral nanostructures harnessing a bespoke few-repeat hybrid multilayer, comprising metallic ([Pt/CoB$(x)$/Ir]$_3$) and oxide (Pt/CoB$(y)$/MgO) stack components, both designed with matched switching fields, and strongly coupled by IEC. 
    The ZF magnetic states of the nanodots are determined by the interplay of dot-size-induced confinement effects and CoB-thickness-driven stack anisotropy. 
    Smaller dots ($W \lesssim 600$ nm) stabilize ZF skyrmions over a substantial range of $K_{\rm eff}$ and dot sizes.  
    The skyrmion size, as small as 50 nm at ZF, can be tuned by up to 2.5$\times$ by engineering the stack anisotropy and geometric parameters.
    Finally, large-$W$, low-$K_{\rm eff}$ dots host higher-order target skyrmions at ZF with chiral rotation up to 3$\pi$, whose stability relative to skyrmionic multiplets can be tailored by small bias fields realizable using stack engineering. 

\paragraph{Impact} 
    The presented stability of ZF skyrmions in our few-repeat hybrid multilayer nanodots has practical consequences for several efforts on chiral spin textures. 
    On one hand, the inherent realization of higher-order skyrmions and their interplay with conventional skyrmions in highly-scalable multilayer nanodots is expected to enable their much-sought functionalization via device concepts harnessing topological transitions \cite{Song.2020}.
    Notably, the stability of conventional ZF skyrmions over a wide range of magnetic and geometric parameters within MTJ-compatible hybrid multilayer nanodots is of critical importance, given their limited stability in prior MTJ-compatible stacks \cite{Boulle.2016, Juge.2018}. 
    Our hybrid stack provides a viable platform for realizing electrical skyrmion readout and manipulation \cite{Shaohai.2023}, and can be generalized to other materials and geometries, notably including racetracks, to achieve all-electrical skyrmionics \cite{Fert.2017, Back.2020}.

\begin{center}\rule[0.5ex]{0.5\columnwidth}{0.5pt}\par\end{center}

\phantomsection
\def\bibsection{\section*{\refname}} 
\linespread{1.00}
\setlength{\parskip}{0.2ex}
\bibliography{Dots2_ref.bib}

\vspace{0.1ex}
\begin{center}
\rule[0.5ex]{0.5\columnwidth}{0.2pt}
\par\end{center}
\vspace{0.1ex}

\textsf{\textbf{\small{}Acknowledgments.}}{\small{}
This work was supported by the SpOT-LITE programme (Grant No. A18A6b0057), funded by Singapore's RIE2020 initiatives, and by NUS funds (Grant No. A-0004544-00-00).}

\end{document}